\documentclass[12pt]{article}

\renewcommand{\baselinestretch}{1.7}

\usepackage{graphicx}
\usepackage{multirow,latexsym,amsfonts,cite}
\usepackage{amsmath}
\usepackage{amssymb}
\usepackage{epsfig, comment}
\usepackage{latexsym}
\usepackage{color}

\newcommand{\be}{\begin{equation}}
\newcommand{\ee}{\end{equation}}
\newcommand{\bea}{\begin{eqnarray}}
\newcommand{\eea}{\end{eqnarray}}

\newcommand{\vect}[1]{\vec{#1}}

\begin{document}
\begin{titlepage}
\null
\begin{flushright}
%hep-th/YYMMXXX
%HIP-2007-62/TH
\end{flushright}

\vskip 1.2cm

\begin{center}
%{\Huge \bf Preliminary version 1, 17.11.2009}

  {\Large \bf   Magnetic Monopole in Noncommutative Space-Time \\
  and Wu-Yang Singularity-Free Gauge Transformations \\
   % {\color{magenta} (Part 1)}
   }

\vskip 1.4cm

{\bf{\large{Miklos L\aa ngvik, Tapio Salminen and Anca Tureanu}}}\\

\vskip .1cm
\normalsize

%  {\large Miklos L\aa ngvik\footnote{miklos.langvik@helsinki.fi} and Tapio Salminen\footnote{tapio.salminen@helsinki.fi}}

\vskip 0.5cm

  {\large \it Department of Physics, University of Helsinki, \\
              P.O. Box 64, FIN-00014 Helsinki, Finland}

\vskip 1cm

{\bf Abstract}

\end{center}

\renewcommand{\baselinestretch}{1.5}\selectfont

\noindent{We investigate the validity of the Dirac Quantization Condition (DQC) for magnetic monopoles
in noncommutative space-time. We use an approach which is based on an extension of the method introduced by Wu and Yang. To study the effects of noncommutativity of space-time, we consider the gauge transformations of $U_\star(1)$ gauge fields and use the corresponding deformed Maxwell's equations. Using a perturbation expansion in the noncommutativity parameter $\theta$, we show that the DQC remains unmodified up to the first order
in the expansion parameter. The result is obtained for a class of noncommutative source terms, which reduce to the Dirac delta function in the commutative limit. }

\vspace*{1cm}

\end{titlepage}

\newpage

\section{Introduction \label{intro}}

Even though magnetic monopoles remain unobserved experimentally, they have attracted interest for decades both experimentally and theoretically.
%For a review on the experimental searches, see \cite{gia}.
In 1931, Dirac \cite{Dirac} showed for the first time that the existence of a magnetic monopole would imply the quantization of electric charge. This, along with the duality-like symmetry of Maxwell's equations, are a major motivation for the study of monopoles. The DQC,
\be
{2ge\over \hbar c} = \mathrm{integer} = N,
\label{Diraccond1}
\ee
is a topological property of space. In 1975, the singular potentials that Dirac's derivation results in, were better understood when Wu and Yang \cite{WuYa} rederived the DQC by a new method based on singularity-free gauge transformations.

In this paper we study the validity of the DQC in noncommutative space-time using the Wu and Yang approach. The motivation for such a study is that since the noncommutativity of space-time is expected to affect the very
short distances, the singularity structure could also be affected.  We consider the noncommutative (NC) Weyl-Moyal space-time defined by the commutator:
\be
[\hat{x}^\mu ,\hat{x}^\nu]=i\theta^{\mu \nu}\\, \label{nccom}
\ee
where $\theta^{\mu\nu}$ is a constant antisymmetric matrix (not a Lorentz tensor).

The study of different types of noncommutative space-time models is mainly motivated by open string theory with a constant background field \cite{SeiWitt} and the attempt to combine quantum mechanics with classical gravity \cite{DopFredRob}. In this work we only consider the so-called space-space noncommutativity, i.e. $\theta^{0i}=0$, because of the issues of unitary \cite{GM} and causality \cite{SST} which appear when time does not commute with space.

The use of the commutator \eqref{nccom} has interesting consequences, most notably the breaking of Lorentz invariance. However, the corresponding noncommutative space-time has a richer symmetry than the subgroup to which the Lorentz group is broken, the so-called twisted Poincar\'e symmetry \cite{ChaKuNiTu}, which has the same representation content as the usual Poincar\'e symmetry.
In addition, UV and IR divergences are mixed in noncommutative field theories \cite{MinRaaSei}. This mixing is the manifestation of the fact that the short and long distance effects in such theories are intimately related.
Thus, it is interesting to
explore the DQC in the noncommutative case, since it probes the singularity structure of the theory.

For noncommutative field theories we shall use the well-known Weyl-Moyal star product and the fact that the gauge group $U_\star(1)$ is
non-Abelian and thus the form of the gauge transformations of fields will differ from the commutative $U(1)$ theory \cite{Haya}.  In this work we shall use the asymptotic expansion of the Weyl-Moyal $\star$-product:
\be
(\phi \, \psi) (x)\rightarrow \left(\phi \star \psi\right)(x) \equiv
\left[\phi(x)e^{\frac{i}{2}\theta_{\mu \nu} \label{star}
\frac{\overleftarrow\partial}{\partial x _\mu}
\frac{\overrightarrow\partial}{\partial y_\nu}}
\psi(y)\right]_{x=y}\,.
\ee
Since we shall be working outside of singularities and with continuous functions we are free to use \eqref{star} which is defined for smooth functions.

The paper is arranged as follows: In section 2 the approach to the DQC of Wu and Yang \cite{WuYa} in commutative space is briefly reviewed. In section 3 we discuss the modifications needed when considering the DQC in noncommutative space-time. In section 4 we study the NC gauge transformations to first order in $\theta$ and in sections 5 and 6 we define and solve the noncommutative Maxwell's equations to first order in $\theta$ and discuss how the DQC can be kept valid to this order. In section 7 we make our concluding remarks.
%{\color{red}why it cannot be obtained in the perturbative approach in second order}.

\section{DQC in the Wu-Yang approach}\label{WuYang}

When describing a magnetic monopole in the Dirac approach \cite{Dirac}, one is led to a singularity in the gauge potential $A_\mu$ for the magnetic field -- the Dirac string.
The string is rotatable by a gauge transformation and thus cannot be observed, but the gauge transformations used for the rotation are also singular. This could be somewhat troubling.
In the approach of Wu and Yang \cite{WuYa}, the singularity problem is circumvented by dividing the whole space into two overlapping hemispheres and by defining a singularity-free potential in each hemisphere.
% the potentials $A_\mu^N$ and $A_\mu^S$ such that each is singularity-free.
In the original paper the space R is divided as:
\bea
R^N: & 0 \leq \theta < \pi/2 + \delta, \hspace{5pt} r > 0, \hspace{5pt} 0 \leq \phi < 2\pi, \hspace{5pt} t \in (-\infty, \infty) \nonumber \,, \\
R^S: & \pi/2 - \delta < \theta \leq \pi, \hspace{5pt} r > 0, \hspace{5pt} 0 \leq \phi < 2\pi, \hspace{5pt} t \in (-\infty, \infty) 
\label{regionsab}
\eea
and the two gauge fields $A_{\mu}^N$ and $A_{\mu}^S$ are taken to be:
\bea\label{potentials}
A_t^N = A_r^N = A_{\theta}^N = 0, & A_{\phi}^N = {g\over r\sin\theta}(1 - \cos\theta), \nonumber \\
A_t^S = A_r^S = A_{\theta}^S = 0, & A_{\phi}^S = -{g\over r\sin\theta}(1 + \cos\theta).
\eea
The conditions the potentials need to satisfy are the following:
\begin{enumerate}
\item{In the overlapping region they are gauge transformable to each other;}
\item{Their curls give the magnetic field;}
\item{Both potentials are singularity-free in their respective regions of validity.}
\end{enumerate}
For the potentials \eqref{potentials}, the gauge transformation is given by
\be
S = S_{ab} = e^{-i\alpha} = e^{{2ige\over \hbar c}\phi}.
\label{gaugetrans}
\ee
This gauge transformation remains single-valued only if the condition
\be
{2ge\over \hbar c} = \mathrm{integer} = N
\label{Diraccond}
\ee
is satisfied. Equation \eqref{Diraccond} is exactly the quantization condition due to Dirac \cite{Dirac}.

\section{Wu-Yang procedure in Moyal space-time \label{repstar}}

To check the validity of the Dirac quantization condition in NC space-time, we shall use a slightly modified version of the Wu-Yang procedure. In commutative space-time, Wu and Yang looked for a gauge transformation from one hemisphere to the other and required that the potentials in each hemisphere give the magnetic field. In NC space-time the situation is modified since the $U_\star(1)$ group is non-Abelian and it is not clear what the magnetic field is in this case.
Therefore we shall look for a potential in each hemisphere, $A_\mu^N(x)$ and $A_\mu^S(x)$, such that:
\begin{enumerate}
\item{The potentials are gauge transformable to each other in the overlap region of the potentials. For the non-Abelian group $U_\star(1)$ this means that we require:
\be \label{gaugetransform}A_{\mu}^{N/S} (x)\rightarrow U(x)\star A_{\mu}^{N/S}(x)\star U^{-1}(x) - iU(x)\star\partial_{\mu}U^{-1}(x)=A_{\mu}^{S/N}(x)\,.\ee}
\item{Both potentials satisfy Maxwell's equations with an appropriate source for the magnetic charge.}
\item{The potentials remain singularity-free in their respective regions of validity, i.e., Maxwell's equations are solved in such a way that noncommutativity does not produce
new singularities into the potentials.}
\end{enumerate}
The analogy between these conditions and the conditions in the commutative case (see section \ref{WuYang}) is apparent.

Requiring these three conditions, we may then consider which types of sources would be compatible with the DQC. This follows because from both equations we are able to solve for the difference in the potentials $A_\mu^N(x)-A_\mu^S(x)$ (perturbatively) and since both equations need to be satisfied simultaneously we get an equation relating the source term contained in Maxwell's equations and the gauge transformation parameter $\lambda(x)$.

We shall treat the problem as a perturbation series to first order in $\theta$. We use a notation where the NC gauge field $A_\mu$ is expanded as $A_\mu = A^0_\mu+A^1_\mu+A^2_\mu+O(\theta^3)$. Here the upper index corresponds to the order in $\theta$. In this notation, the gauge transformation parameter is (symbolically) expanded as $\lambda =\lambda^0+\lambda^1+ \lambda^2+O(\theta^3)$. To preserve the DQC we require that the $\theta$-corrections to $\lambda$ can be put to zero (or a constant), i.e. $\lambda =\lambda^0+C$, while satisfying the three above requirements. We begin by calculating the finite gauge transformations of the fields to first order in $\theta$.

\section{Noncommutative gauge transformations}\label{NCGT}

The noncommutative gauge transformations under the group $U_\star(1)$, with group elements $$U^{-1}(x) = e_{\star}^{i\lambda(x)} = 1 + i\lambda(x) + {i^2\over 2!}\lambda(x)\star\lambda(x) + \dots,$$ are given by
\be
A_{\mu}(x) \rightarrow U(x)\star A_{\mu}(x)\star U^{-1}(x) - iU(x)\star\partial_{\mu}U^{-1}(x). \label{GT}
\ee
There will be $\theta$-contributions from the non-trivial gauge group element $U(x)$, as well as from the $\star$-products between the factors of \eqref{GT}.
% We start by calculating the gauge group element to second order in $\theta$.

%For this to be useful for our purpose, we must compute the finite gauge transformations to second order in $\theta$. We do this by first
%calculating the gauge group elements $U^{-1}(x)$ to second order in $\theta$.
Using the result
\be
\theta^{ij}\partial_i\lambda(x)\partial_j\lambda(x) = 0\,, \label{stuff}
\ee
%We can then compute the term $U^{-1}(x)$ to first order. It is given by
we see that the gauge group element in first order in $\theta$ remains unmodified:
\bea
e_{\star}^{i\lambda(x)} & = & 1 + i\lambda(x) + {i^2\over 2!}\lambda(x)\star\lambda(x) + {i^3\over 3!}\lambda(x)\star\lambda(x)\star\lambda(x) + \dots \nonumber \\
& = & e^{i\lambda(x)} + \mathcal{O}(\theta^2) \label{final}.
\eea

Next, we go back to \eqref{GT} and calculate the full expression for the gauge transformation in first order. Writing the noncommutative field $A_i(x)$ as $A_i(x) = A_i^0(x)+A_i^1(x)+A_i^2(x)+ \dots,$
%where $A_i^0(x)$ is the term of zeroth order in $\theta$, $A_i^1(x)$ is the term of first order in $\theta$, et.c.,
we can calculate the gauge transformation \eqref{GT} to
first order in $\theta$. It is given by
\bea \label{gaugetransexpansion}
A_i(x) & \rightarrow & e^{-i\lambda}\star(A_i^0(x) + A_i^1(x))\star e^{i\lambda} - ie^{-i\lambda}\star\partial_ie^{i\lambda} + \mathcal{O}(\theta^2) \nonumber \\
& = & A_i^0(x) + A_i^1(x) + \theta^{kl}\partial_k\lambda\partial_lA_i^0(x) + \partial_i\lambda + {\theta^{kl}\over 2}\partial_k\lambda\partial_l\partial_i\lambda + \mathcal{O}(\theta^2) \label{a:s}.
\eea
From this we have the following gauge transformations in the zeroth and first order in $\theta$:%. To first order we have
\bea
A_i^0(x) & \rightarrow & A_i^0(x) + \partial_i\lambda \label{crap23} \,,\\
A_i^1(x) & \rightarrow & A_i^1(x) +  \theta^{kl}\partial_k\lambda\partial_lA_i^0(x) + {\theta^{kl}\over 2}\partial_k\lambda\partial_l\partial_i\lambda. \label{crap24}
\eea

To conclude, due to the first requirement in section \ref{repstar}, we require up to first order that the following equations hold:
\bea
A_i^{N_0}(x) & = &A_i^{S_0}(x) + \partial_i\lambda  \label{1stord}\,,\\
A_i^{N_1}(x) & = &A_i^{S_1}(x) +  \theta^{kl}\partial_k\lambda\partial_lA_i^{S_0}(x) + {\theta^{kl}\over 2}\partial_k\lambda\partial_l\partial_i\lambda \,.\label{2ndord}
\eea
Next we shall move on to consider the second requirement of section \ref{repstar}, i.e. that the potentials satisfy Maxwell's equations.

\section{Noncommutative Maxwell's equations in first order \label{1st}}

In Weyl-Moyal space Maxwell's equations for a static monopole are:
%The noncommutative Maxwell's equations for the dual field strength tensor $\mathcal{F}_{\mu\nu}= {1\over 2}\epsilon^{\mu\nu\gamma\delta}F_{\gamma\delta}$ will be taken to be (for $U_{\star}(1)$)
\begin{align}
\epsilon^{\mu\nu\gamma\delta} D_\nu\star\mathcal{F}_{\gamma\delta} &= 0  \label{theother}\,, \\
D_\mu\star\mathcal{F}^{\mu\nu} &= J^\nu \,, \label{sourceeqn}
\end{align}
where $\mathcal{F}_{\mu\nu}= {1\over 2}\epsilon^{\mu\nu\gamma\delta}F_{\gamma\delta}$ is the dual field strength tensor. The NC $U_\star(1)$ field strength tensor and the covariant derivative are given by
\begin{align}
F_{\mu\nu} &= \partial_\mu A_\nu - \partial_\nu A_\mu - ie[A_\mu , A_\nu]_{\star}\,, \\
D_\nu &= \partial_\nu - ie[A_\nu,\cdot]_{\star} \,.
\end{align}
We shall look at the equations \eqref{theother} and \eqref{sourceeqn} as a perturbative series in $\theta$ and check whether we can find solutions for them perturbatively. For the source we have $J^i=0$ and $J^0 \equiv \rho(r) = 4\pi g\delta(r) + \rho^{1}(r) + \rho^2(r) + \mathcal{O}(\theta^3)$, where the superscript denotes the order of $\theta$. In this way the total noncommutative magnetic charge is defined as
\be
g_{NC} = \int J^0(x)d^3x
\ee
and it is therefore
gauge invariant. Observe that $g_{NC}$ becomes a perturbation series in $\theta$ and consequently $g$ from the commutative Maxwell's monopole equations becomes a coupling constant which coincides with the definition for magnetic charge in the commutative limit.

We should mention that we have the additional consistency condition for eq. \eqref{sourceeqn},
\be
D_{\mu}\star J^{\mu} = 0.
\ee
We do, however, not need to consider this condition separately because the source and monopole equations are static, $J^i = 0$, and all electric fields are set to zero (i.e. $A_0 = 0$, in the static case).

Because of the static case we consider, the other two Maxwell's equations contained in  \eqref{theother} and \eqref{sourceeqn}
are identically satisfied, and consequently we shall refer to equation \eqref{theother} as Amp\`ere's law and to equation \eqref{sourceeqn} as Gauss's law. In the following, we shall use units in which $\hbar = c = g = e = 1$ throughout,
but the units will be restored whenever we return to discussing the DQC.

The gauge covariant form of the equations does make one worry about the existence of gauge invariant noncommutative electric and magnetic fields. This problem can be
overcome if we take the view that the electric and magnetic fields are gauge invariant combinations of the potentials. That is, if we can find gauge invariant combinations of
the noncommutative potential that reduce to the electric and magnetic fields in the $\theta \rightarrow 0$ limit, it is very well justified to call these combinations
the noncommutative electric and magnetic field respectively. These combinations can be found by the use of Wilson lines as in \cite{Gross2}, where gauge invariant operators
in noncommutative gauge theory were constructed. These operators are given in momentum space, but by a usual, \emph{commutative} inverse Fourier transformation they can be transformed back
to coordinate space. Using that result, one may define a gauge invariant object constructed from the $U_{\star}(1)$ field strength tensor $F^{\mu\nu}$ as:
\be
G^{\mu\nu} = \int d^4ke^{-ikx}\Big[\int d^4 x  F^{\mu\nu} \star W(x,C) \star e^{ikx}\Big] \,,\label{GIObs}
\ee
where $W(x,C)$ is the noncommutative $U_{\star}(1)$ Wilson line:
\be
W(x,C)= P_{\star} \exp \left(i g \int_0^1 d \sigma {d \zeta ^{\mu} \over
d \sigma} A_{\mu}(x+\zeta(\sigma))\right),
\ee
and where $C$ is the curve which is parameterized by $\zeta^{\mu}(\sigma)$ with
$0\leq \sigma \leq 1$, $\zeta(0) = 0$. $\zeta(1)=l$ and satisfies the condition $l^{\nu}=k_\mu \theta^{\mu\nu}$,  $l$ being the length of the curve.  $P_{\star}$ denotes
path ordering with respect to the star product:
\be
W(x,C)= \sum_{n=0}^{\infty} (ig)^n
\int_0^1 d\sigma_1 \int_{\sigma_1}^1 d \sigma_2 ...
\int_{\sigma_{n-1}}^1 \!\!\!\!\!\! d \sigma_n \
\zeta^{'}_{\mu_1}(\sigma_1) ... \zeta^{'}_{\mu_n}(\sigma_n)
A_{\mu_1}(x+\zeta(\sigma_1))\star ... \star A_{\mu_{n}}(x+\zeta(\sigma_n)).
\ee
%Equation \eqref{GIObs} may look somewhat asymetric, but that does not matter.
Equation \eqref{GIObs} is a gauge invariant combination of the noncommutative potential, that reduces to the
commutative field strength in the limit $\theta \rightarrow 0$. Therefore the $F^{0i}$ and $F^{ij}$ parts of the noncommutative field strength may be attributed
to the noncommutative electric and magnetic fields, such that $G^{0i}$ is the noncommutative electric field and $\epsilon_{ijk}G^{jk}$ is the noncommutative magnetic field.

One should point
out that the shape of the curve $C$ gives rise to different gauge invariant objects, and therefore the definition of the magnetic and electric fields in \eqref{GIObs} is ambiguous. It may
be that straight Wilson lines are the best choices as then the point of attachment of $F^{\mu\nu}$ to the Wilson line does not matter as argued in \cite{Gross2}. However, the definitions of the
gauge invariant fields are only given here for a better understanding of the noncommutative Maxwell's equations and as we shall not need them in the following, we do not discuss this ambiguity
further.

\noindent{\large {\bf Amp\`{e}re's law}: \normalsize We start by investigating \eqref{theother}}:
\be
\epsilon^{\mu\nu\gamma\delta} D_\nu\star\mathcal{F}_{\gamma\delta} = \frac 12 \epsilon^{\mu\nu\gamma\delta}\epsilon_{\gamma\delta\alpha\beta} D_\nu\star F^{\alpha\beta} = 2 D_\nu\star F^{\mu\nu} = 0 \,,
\ee
in the first order in $\theta$. Since the electric field is set to zero, $F^{0i}=E^i=0$, and all time derivatives vanish, the indices $\mu$ and $\nu$ run over the spatial coordinates only. Consequently, we have
\begin{align}
D_k\star F^{ik} =& \partial_k (\partial^i A^k_1-\partial^k A^i_1) + \partial_k (\partial^i A^k_0-\partial^k A^i_0)\notag\\
&-i\partial_k[A^i_0 , A^k_0]_\star -i [A_k^0,\partial^i A^k_0-\partial^k A^i_0-i[A^i_0 , A^k_0]_\star]_\star + \mathcal{O}(\theta^2) \notag\\
=& \epsilon^{ikp} \partial_k B^1_p+\epsilon^{ikp} \partial_k B^0_p
%&=\partial_\nu (\partial^\mu A^\nu_1-\partial^\nu A^\mu_1) + \partial_\nu (\partial^\mu A^\nu_0-\partial^\nu A^\mu_0)
+\theta^{pq} \{\partial_k(\partial_p A^i_0 \partial_q A^k_0) +\partial_p A_k^0 \epsilon^{ikp}\partial_q B^0_p \}+ \mathcal{O}(\theta^2)\,,
%\rightarrow - D_\nu\star F^{\mu\nu}(A^1) =  \partial_\nu F^{\mu\nu} (A^0) - -\partial
\end{align}
where we have denoted $(\partial^i A^k_1-\partial^k A^i_1)$ by $\epsilon^{ikp} B^1_p$.
% However, because the time derivatives vanish and the electric field is set to zero,
Hence we find Amp\`{e}re's law up to first order in $\theta$ to have the form:
\begin{align}
(\nabla \times \vect{B}^0)^i &=0\,,\notag\\
%&\partial_\nu (\partial^\mu A^\nu_1-\partial^\nu A^\mu_1) + \partial_\nu (\partial^\mu A^\nu_0-\partial^\nu A^\mu_0) = \epsilon^{\mu\nu\gamma} \partial_\nu %B^1_\gamma(A^1)+\epsilon^{\mu\nu\gamma} \partial_\nu B^0_\gamma(A^0) \\
(\nabla \times \vect{B}^1)^i &= -\theta^{\gamma\delta} \left[ \partial_j(\partial_\gamma A_0^i \partial_\delta A_0^j) + \partial_\gamma A_j^0 \epsilon^{ijk}\partial_\delta B^0_k   \right] \,, \label{amper}
\end{align}
where
% we have used $\nabla \times B^0 = 0 $ and
$i,j,k = 1,2,3$.

\vspace{10pt}

\noindent{\large {\bf Gauss's law:} \normalsize Next we shall do the same for Gauss's law:
\begin{align}
D_\mu \star \mathcal{F}^{\mu\nu} &= \frac 12  \epsilon^{\mu\nu\alpha\beta}D_\mu \star F_{\alpha\beta} = J^\nu \,.
\end{align}
Since $J^i=0$, we set $\nu=0$ and as in the above the indices run over the spatial coordinates only.
%In the following we shall denote the 0th and 1st order terms of the source by $J^{\nu^0}$ and $J^{\nu^1}$, respectively ($J^{\nu^2}$ also appears later).
Separating again the relevant term
$\nabla \cdot \vect{B_1}$ and using $$\frac 12  \epsilon^{i 0 jk} \partial_i (\partial_j A^0_k- \partial_k A^0_j)=-\frac 12  \epsilon^{ijk} \partial_i (\partial_j A^0_k- \partial_k A^0_j)=-\nabla\cdot \vect{B_0} = 4\pi\delta(r),$$ we obtain:
\begin{align}
D_i \star \mathcal{F}^{i0} - 4\pi\delta(r) =&\, \frac 12  \epsilon^{i0jk} \partial_i (\partial_j A^0_k- \partial_k A^0_j)- 4\pi\delta(r)+\frac 12  \epsilon^{i0jk} (-i)[A_i, \partial_j A^0_k- \partial_k A^0_j]_\star \notag \\
 &+\frac 12\epsilon^{i0jk} \partial_i (-i)[A_j^0,A_k^0 ]_\star +\frac 12  \epsilon^{i0jk} \partial_i (\partial_j A^1_k- \partial_k A^1_j) - \rho^1(x) \notag\\
=&\, i \frac 12 \epsilon^{ijk} \left( [A_i, \partial_j A^0_k- \partial_k A^0_j]_\star+ [\partial_i A_j^0,A_k^0 ]_\star+[A_j^0,\partial_i A_k^0 ]_\star \right)
 - \frac 12  \epsilon^{ijk} \partial_i \epsilon_{jkl} B_1^l-\rho^1(x)  \notag \\
 =&-\frac 12  \epsilon^{ijk} \partial_i\epsilon_{jkl} B_1^l - \rho^1(x) \,.
\end{align}
%Using the fact that $J^i = 0 $ we get the simple result:
Since $-\frac 12  \epsilon^{ijk} \partial_i\epsilon_{jkl} B_1^l=\nabla \cdot \vect{B}_1 $, we find the simple result:
\begin{align}
%-\frac 12  \epsilon^{\mu0\alpha\beta} \partial_\mu \epsilon_{\alpha\beta\gamma} B_1^\gamma =\frac 12  \epsilon^{\mu\alpha\beta} \partial_\mu %\epsilon_{\alpha\beta\gamma} B_1^\gamma =
\nabla \cdot \vect{B}_0 &= -4\pi\delta(r)\notag\,,\\
\nabla \cdot \vect{B}_1 &= -\rho^1(x) \,.
\label{gaush}
\end{align}

\vspace{10pt}

\noindent{\large {\bf Combining Amp\`{e}re's and Gauss's laws:}} \normalsize We combine the equations of motion of Gauss \eqref{gaush} and Amp\`ere \eqref{amper} in the usual way, with the help of the identity from vector calculus $\nabla^2 \vec{B} = \nabla(\nabla \cdot \vec{B})+\nabla \times (\nabla \times \vec{B}) $.
Since the form of Gauss's law is so simple, we obtain:
\begin{align}
(\nabla^2 B_1(A_1))^i &= -\partial^i \rho^1-\theta^{pq}\epsilon^{ijk} \partial^j  \left[ \partial^l(\partial^p A_0^k \partial^q A_0^l) + \partial^p A^l_0 \epsilon^{klm}\partial^q B_0^m   \right] \notag \\
&=-\partial^i\rho^1-\theta^{pq} \{ \epsilon^{ijk} \partial^j \partial^l(\partial^p A_0^k \partial^q A_0^l) + \partial^m (\partial^p A^i_0 \partial^q B_0^m )- \partial^m (\partial^p A^m_0 \partial^q B_0^i ) \} \notag \\
&=-\partial^i \rho^1-\theta^{pq} \{\epsilon^{ijk} \partial^l(\partial^p A_0^k  \partial^j\partial^q A_0^l)- 2\partial^m (\partial^p A^m_0 \partial^q B_0^i ) - \partial^m (\partial^p B_0^m \partial^q A^i_0) \}\,. \label{lapb1}
\end{align}

\section{Solution of the noncommutative Maxwell's equations \label{crud}}

To complete criterion 2 for our potentials in section \ref{repstar}, we need to solve equation \eqref{lapb1}. We do this by choosing a frame of reference, i.e. fix $\theta^{pq}$. We will first choose $\theta^{12}=-\theta^{21}$, whilst all other components of $\theta$ are set to zero. Furthermore, we use the original potentials of Wu and Yang \eqref{potentials} in cartesian coordinates:
\begin{align}
A_1^{N_0}  = & {-y (r - z)\over (x^2 + y^2) r},\quad A_2^{N_0}  =  {x (r - z)\over (x^2 + y^2) r},\quad A_1^{S_0}  =  {y (r + z)\over (x^2 + y^2) r},\quad \;A_2^{S_0}  =  {-x (r + z)\over (x^2 + y^2) r}\,,\label{potstuff}\\
& \hspace{2.5cm} A_3^{N_0}  =  A_3^{S_0} = A_0^{N_0} = A_0^{S_0} = 0. \nonumber
\end{align}
Here $N_0$ and $S_0$ denote the zeroth order terms in $\theta$ in the northern and southern hemispheres, respectively, and $r=\sqrt{x^2+y^2+z^2}$.

Since the potentials \eqref{potstuff} are only defined outside the origin, the expression \eqref{lapb1} contains two problematic terms. For all components $i$, there is a term $\theta^{pq}(\partial_p \nabla \cdot \vect{B}_0)\partial_q A^i_0$ and when $i=3$ we have the additional term $\theta(\nabla \cdot \vect{B}_0)^2$.
These are somewhat ambiguous in our construction since for the Maxwell's equations the origin is included but for 
the zeroth order potentials $A^i_0$, this point is removed. This implies that once we expand the noncommutative field as $A^i_{NC} = A^i_0 + A^i_1 + A^i_2 + ...$,
the origin is not included in the range of values that $A^i_{NC}$ can take. Consequently, we can ignore any contribution to the total noncommutative field $A^i_{NC}$ that contributes only at the origin, making the 
theory more singular.
% Such terms are singular at the origin of the coordinates, where the monopole is, but since the equations for the potentials are supposed to hold everywhere, with the exception of the origin, the problematic terms can be discarded. 
Therefore, terms such as $\nabla^2 B_i^1 = \theta^{pq}(\partial_p \delta^3(r))\partial_q A^i_0$ can be considered to
contribute to $B_i^1$ only at $r=0$ and consequently ignored. The same goes for terms of the form $\theta( \delta^3(r))^2$.
In Maxwell's equations we need to keep these terms, for the consistency of the perturbative approach.

By using
the components \eqref{potstuff} and \eqref{lapb1} for $\theta = \theta^{12}$, we get the following Laplace equations for the difference in $\vec{B}_1$ in the overlap of the
two potentials:
\begin{align}
\nabla^2(B^{N_1} - B^{S_1})_1 & =  {12\theta x(3r^2 - 5z^2)\over (r^2-z^2)r^7} - \partial_1\rho^{N_1} + \partial_1\rho^{S_1} \label{b1eq} \,,\\
\nabla^2(B^{N_1} - B^{S_1})_2 & =  {12\theta y(3r^2 - 5z^2)\over (r^2-z^2)r^7} - \partial_2\rho^{N_1} + \partial_2\rho^{S_1} \label{b2eq} \,,\\
\nabla^2(B^{N_1} - B^{S_1})_3 & =  {4\theta z(45(r^2 - z^2)^3 + 70(r^2 - z^2)^2z^2 + 56(r^2-z^2)z^4 + 16z^6)\over (r^2-z^2)^3r^7} - \partial_3\rho^{N_1} + \partial_3\rho^{S_1}.  \label{b3eq}
\end{align}
In these equations the vector index $0$ for the source term from equation \eqref{gaush} has been dropped. The superscript $N_1$ in $\rho^{N_1}$ means the northern hemisphere and the first order noncommutative correction to
the source; similarly, $S$ denotes the southern hemisphere.
%The notation for the southern hemisphere is denoted by an $S$.

The homogenous part ($\rho^{N_1}=\rho^{S_1}=0$) of these equations is solved by:
% These equations, neglecting the source, are solved by:
\begin{align}
(B^{N_1} - B^{S_1})_1 &= {2\theta x(2r^2-3z^2)\over (r^2-z^2)r^5}\,, \label{soln1} \\
(B^{N_1} - B^{S_1})_2 &= {2\theta y(2r^2-3z^2)\over (r^2-z^2)r^5}\,, \label{soln2} \\
(B^{N_1} - B^{S_1})_3 &= {2\theta z(6(r^2-z^2)^2 + 5(r^2-z^2)z^2 + 2z^4)\over (r^2-z^2)^2r^5}\,.\label{soln3}
\end{align}
%In equations \eqref{soln1}, \eqref{soln2} and \eqref{soln3} we have not yet taken into account the contribution from the source term.
The solutions to these equations are given by:
\bea
A^{N_1}_1 - A^{S_1}_1 & = & {2\theta yz(2r^2-z^2)\over (r^2-z^2)^2r^3} \label{solnA1} \,,\\
A^{N_1}_2 - A^{S_1}_2 & = & -{2\theta xz(2r^2-z^2)\over (r^2-z^2)^2r^3} \label{solnA2}\,,\\
A^{N_1}_3 - A^{S_1}_3 & = & 0\,. \label{solnA3}
\eea
Now we would like to compare these equations with those coming from the gauge transformation \eqref{gaugetransform}. That is, we try to satisfy
criterions 1 and 2 from section \ref{repstar} simultaneously. More specifically, to first order in $\theta$ we use the explicit formula \eqref{2ndord}:
%To make a comparison with  \eqref{solnA1}, \eqref{solnA2} and \eqref{solnA3} our next step is to see what the gauge transformation \eqref{2ndord} gives using the potentials \eqref{potstuff}.
\be
A_i^{N_1}(x) - A_i^{S_1}(x) =  \theta\left(\partial_1\lambda\partial_2A_i^{S_0}(x) - \partial_2\lambda\partial_1A_i^{S_0}(x)\right) + {\theta\over 2}(\partial_1\lambda\partial_2\partial_i\lambda -
\partial_2\lambda\partial_1\partial_i\lambda).
\ee
Inserting the potentials of Wu and Yang \eqref{potstuff} and $\lambda=\lambda_0+\mathcal{O}(\theta^2)=\frac{2ge}{\hslash c }\phi+\mathcal{O}(\theta^2)$, where $\phi = \arctan\left({y\over x}\right)$, we recover exactly the equations \eqref{solnA1}, \eqref{solnA2} and \eqref{solnA3}.
In other words, there exist potentials $A_\mu^{N_1}$ and $A_\mu^{S_1}$ that are gauge transformable to each other and satisfy the equations of motion
 as long as the
first order contribution to the source term does not change the solution of the equations \eqref{soln1}, \eqref{soln2} and \eqref{soln3}. As these can be solved with the Green's function for the Laplace equation, we have the following condition:
\be
B^{N_1(source)}_i - B^{S_1(source)}_i = -\int{\partial'_i(\rho^{N_1}(r')-\rho^{S_1}(r'))d^3r' \over \mid r'-r\mid} = 0\,. \label{slpp}
\ee

 The symmetries of the equations will further constrain the form of a possible source term. It turns out that the ordinary delta-function source is not compatible with these symmetries and thus needs to be modified. This will be discussed below, in subsection $\ref{creed}$. First, we will check that our results are not sensitive to the choice
%of coordinates, i.e. to the choice
of the noncommutative plane.
%for the first order term of the noncommutative source.

Since we chose our noncommutative plane to be parallel with the overlap, i.e. the $(x,y)$-plane, we should check the above result for another choice of $\theta$
% is parallel to the plane of the overlap of the two potentials $A^{N}_i$ and $A^S_i$ and
as the results might differ when the two planes are not parallel. % We should also check all of this for another choice of Moyal plane,
 % which plane the noncommutativity sits w.r.t. to the overlap of the two potentials.
We should therefore do the same calculation for either $\theta^{13}$ or $\theta^{23}$. Which one of these does not matter because they are symmetric with respect to the plane of overlap. We choose $\theta^{13} = \theta'$.

We begin the analysis in $\theta'$ to first order from equations \eqref{lapb1}. With this choice of $\theta'$, the equations become
\bea
\nabla^2(B^{N_1} - B^{S_1})_1 & = &   {12\theta'xyz(3r^2 - 5z^2)\over (r^2-z^2)^2r^7} - \partial_1\rho^{N_1} + \partial_\rho^{S_1} \label{t13nab1} \,,\\
\nabla^2(B^{N_1} - B^{S_1})_2 & = &  {12\theta'z(-x^2(3r^2 - 5z^2)+(r^2 - z^2)(4r^2-5z^2))\over (r^2-z^2)^2r^7} - \partial_2\rho^{N_1} + \partial_2\rho^{S_1} \label{t13nab2} \,,\\
\nabla^2(B^{N_1} - B^{S_1})_3 & = &  -{12\theta'y(3r^2 - 5z^2)\over (r^2-z^2)r^7} - \partial_3\rho^{N_1} + \partial_3\rho^{S_1}, \label{t13nab3}
\eea
 where we have used the potentials \eqref{potstuff}. The solutions, again neglecting the source, to these equations are
 \bea
 (B^{N_1} - B^{S_1})_1 & = &  {6\theta'xyz\over (r^2 - z^2)r^5}, \label{solnt131} \\
 (B^{N_1} - B^{S_1})_2 & = &  {6\theta'y^2z\over (r^2 - z^2)r^5}, \label{solnt132} \\
 (B^{N_1} - B^{S_1})_3 & = &  {2\theta'y(-2r^2+3z^2)\over (r^2-z^2)r^5}. \label{solnt133}
 \eea
The potentials in the overlap can then be chosen as
\bea
A^{N_1}_1 - A^{S_1}_1 & = & -{2\theta'y^2\over (r^2-z^2)r^3} \label{solnA1t13} \,,\\
A^{N_1}_2 - A^{S_1}_2 & = & {2\theta'xy\over (r^2-z^2)r^3} \label{solnA2t13}\,,\\
A^{N_1}_3 - A^{S_1}_3 & = & 0. \label{solnA3t13}
\eea
These potentials are exactly the same as when we calculate \eqref{2ndord} for the $\theta' = \theta^{13}$-case, and we can conclude in the same fashion as for $\theta^{12}$ that
the DQC holds to first order in $\theta^{13}$, provided condition \eqref{slpp} is satisfied and the correction $\rho^1(r)$ is gauge covariant.

We should make a brief comment on the uniqueness of the above solutions. The solutions are not unique, as we can always add a gradient term $\partial_if(x,y,z)$ to them without changing the equations of motion \eqref{b1eq}, \eqref{b2eq} and \eqref{b3eq}.
However, as we are looking for potentials satisfying the equations of motion and transforming in the right manner under gauge transformations such that the DQC remains unmodified, we are free to choose $\partial_if(x,y,z)$ as we wish, and indeed need to take $\partial_if(x,y,z)=0 $ to preserve the DQC. Therefore we choose $\partial_if(x,y,z)=0 $ and consider the potential differences given by \eqref{solnA1}, \eqref{solnA2} and \eqref{solnA3}.

 At this point we can conclude that there exist potentials $A_\mu^{N_1}$ and $A_\mu^{S_1}$ that are gauge transformable to each other and satisfy the equations of motion as long as the first order contribution to the source term does not change the solution of the equations of motion in the overlapping region in the first order of $\theta$. As was already mentioned above, the symmetries of the equations constrain the source term further and as will be shown in the next subsection we cannot use the ordinary $\delta^3(r)$-function source.

\subsection{Noncommutative corrections to the source term \label{creed}}

In this subsection we consider in general which types of sources are possible  in order to retain the DQC and have
%consistent
noncommutative Maxwell's equations consistent with their gauge symmetry.

The requirement that the first order correction to the source should not affect the solutions of the equations of motion (condition \eqref{slpp}) constrains the form of the source but is not stringent enough to forbid a correction term entirely. Also, equation \eqref{sourceeqn} transforms as $U(x)\star D_\mu\star\mathcal{F}^{\mu0}\star U^{-1}(x)$ on the left-hand side. Namely, it is gauge covariant. Therefore, the source must also transform this way.
Moreover, the left-hand side in equation \eqref{sourceeqn} is $O(1,1)\times SO(2)$ symmetric and consequently, the source must also be that. We shall also, as a correspondence principle, require that we recover the Dirac delta-function for the source when $\theta \rightarrow 0$.

A possible source up to first order in $\theta$, satisfying all the symmetry requirements, is
\be
J^0 =4\pi g \left(\delta^3(r) +\frac{ie}{2\hbar c}\theta^{kl}\partial_k \left(A_{l}\delta^3(r)\right)  \right) +\mathcal{O}(\theta^2)\,. \label{crap}
\ee
This source was found in \cite{dqc} where it was shown that within a specific
quantum mechanical model for noncommutative space-time, the DQC holds to first order in $\theta$. Since the first order source term needs to be a scalar, the above first order part of the source is unique up to a change in the position of the derivative
$\partial_k$ and a change of the constant factor. Since the source is proportional to a delta function $\delta^3(r)$ and derivatives of it, it is easy to convince oneself that its only contribution in first order of the perturbation is to make Maxwell's equations
gauge covariant.

\subsection{Singularity-free potentials}

We should still convince ourselves that the potentials remain singularity-free when the first order corrections are included.
%Furthermore, our next goal will be to repeat the same analysis to second order in $\theta$. This will be presented in \cite{usinsecond}, a work that will be released soon.
To really speak of singularity-free gauge potentials
we need to solve for the potentials $A^{N_1}$ and $A^{S_1}$ (not just their difference) in first order.

If we choose $\theta = \theta^{12}$ as our Moyal plane, we obtain the equations for $B^{N_1}_i$ from \eqref{lapb1} as
\begin{align}
\nabla^2 B^{N_1}_1 & =  {6\theta x(3(r^2-z^2)^2 +(r^2-z^2)z(z-5r) + 2z^3(r-z))\over (r^2-z^2)r^9} \label{BN_1tot} \,,\\
\nabla^2 B^{N_1}_2 & =  {6\theta y(3(r^2-z^2)^2 +(r^2-z^2)z(z-5r) + 2z^3(r-z))\over (r^2-z^2)r^9} \label{BN_2tot} \,,\\
\nabla^2 B^{N_1}_3 & =  2\theta \Big(-{16 \over (r^2-z^2)^3}+{21(r^2-z^2)\over r^8}-{13\over r^6}+{z\over r}\big[{15\over r^6}+{6\over (r^2-z^2)r^4}+{8\over (r^2-z^2)^2r^2} + {16\over (r^2-z^2)^3}\big]\Big) \label{BN_3tot}\,.
\end{align}
%At this point we shall omit the possible correction to the vector potential $\vec{A}^{N_1}_{source}$. It will
%appear again at the end, in the full expression for the potential.

The solutions to the previous equations are given by
\bea
B^{N_1}_1 & = & {\theta x (2 (r^2-z^2)^2 + (r^2-z^2) z (z-6r) + z^3 (r-z))\over (r^2-z^2) r^7} \label{B1}\,,\\
B^{N_1}_2 & = & {\theta y (2 (r^2-z^2)^2 + (r^2-z^2) z (z-6r) + z^3 (r-z))\over (r^2-z^2) r^7}  \label{B2}\,,\\
B^{N_1}_3 & = & \theta \Big(-{2\over (r^2 - z^2)^2} - {7 z^2\over r^6} + {5\over 2r^4} + {6 (r^2 - z^2)^2 z + 5 (r^2 - z^2) z^3 + 2 z^5)\over (r^2 - z^2)^2 r^5}\Big)\,.\label{B3}
\eea
We then proceed to integrate out the potentials for $B^{N_1}_i$. One choice, not introducing new singularities according to requirement 3 in section \ref{repstar}, is e.g.
\bea
A^{N_1}_1 & = & \theta\Big({-2x\arctan({x\over y})\over (r^2-z^2)^2}+{y\over 4}\Big[{7\over r^4} - {2\over (r^2-z^2)r^2} + {4z(2r^2-z^2)\over (r^2-z^2)^2r^3}\Big]\Big) \label{a1whole1} \,,\\
A^{N_1}_2 & = & -\theta\Big({2y\arctan({x\over y})\over (r^2-z^2)^2}+{x\over 4}\Big[{7\over r^4} - {2\over (r^2-z^2)r^2} + {4z(2r^2-z^2)\over (r^2-z^2)^2r^3}\Big]\Big) \label{a1whole2} \,,\\
A^{N_1}_3 & = & 0. \label{a1whole3}
\eea
 From these potentials it is straightforward to obtain the expression for $A^{S_1}_i$, using \eqref{solnA1}, \eqref{solnA2} and \eqref{solnA3}.
It is clear that the potentials are singularity-free on the manifold $\mathbb{R}^3\setminus\{0\}$. We have thus found potentials that remain singularity-free over the manifold $\mathbb{R}^3\setminus\{0\}$ and our
 construction is complete.
 We may note that from the form of equations \eqref{B1}-\eqref{B3} it is clear that the large $r$ limit gives us the commutative theory when we consider only the leading order terms.

\section{Discussion and conclusions \label{conclrem}}

In this paper we have introduced a modified version of the method of Wu and Yang to accomodate noncommutativity of space-time when considering magnetic monopoles. The method is based on perturbation theory with the expansion parameter $\theta$. Using this method we have studied the Dirac quantization condition (DQC) to first order in $\theta$ and found that the condition remains unmodified for a class of sources, that reduce to the Dirac delta function in the commutative limit. Our result serves to clarify the relation between Maxwell's equations and the quantum mechanical model used in \cite{dqc}.

There have been many interesting studies devoted to noncommutative BPS-
monopoles (see \cite{U2, U1, nonpert, lecht} for a non-exhaustive list of references). The works include 
perturbative studies of the $U_{\star}(2)$  \cite{U2} and $U_{\star}(1)$  \cite{U1} BPS-monopoles, as well as 
nonperturbative
studies of the $U_{\star}(1)$ \cite{nonpert} BPS-monopoles, generalized to other groups in \cite{lecht}. These 
constructions share the assumption that the definition  of magnetic charge in the 
BPS-limit
may be taken over, without change,  to the noncommutative case. The legitimacy of this 
assumption  is still an open question,  since the BPS-gauge field should reduce to the 
solution of noncommutative Maxwell's equations, with a magnetic monopole, in order to 
justify the
very name magnetic charge in this context. It should be stressed that,
with a constant $\theta$-matrix, the noncommutativity is  present  everywhere in space 
and thus  the  effect of noncommutativity  can not be assumed to vanish even 
asymptotically far away from the monopole. Although the BPS-constructions \cite{U2, U1, nonpert, lecht} do have 
a topological charge,  it is not
necessarily the same as the noncommutative magnetic charge considered in the present work.

%Before this work there has been a lot of interest devoted to noncommutative BPS-monopoles (see  for a non-exhaustive list of references). The works
%include perturbative studies of the $U_{\star}(2)$ \cite{U2} and $U_{\star}(1)$ \cite{U1} BPS-monopoles as well as nonperturbative studies of the $U_{\star}(1)$ \cite{nonpert} BPS-monopoles, %generalized to other groups in \cite{lecht}.    
%These constructions share the assumption that the definition for magnetic charge in the BPS-limit can without change be taken over to the noncommutative case. This may be too much of an %assumption, as we should have the gauge field reduce to the solution of the \emph{noncommutative} Maxwell's equations with a magnetic monopole in order to be able to justify the name %magnetic charge in the BPS-limit for a noncommutative theory. This follows because with a constant $\theta$-matrix the noncommutativity is present everywhere in space-time and thus cannot be %assumed to vanish very far away from the monopole. It should be stressed that the BPS-constructions prior to this work do have a topological charge, but it is not necessarily the same as the %noncommutative magnetic charge. 

To find out what the noncommutative magnetic charge is, one must first solve Maxwell's equations in noncommutative space for a magnetic monopole to know how the noncommutative magnetic monopole solution looks in order to speak of a magnetic monopole within noncommutative field theory. The aim of this work has therefore been to begin to fill in this gap. The analysis to second order in $\theta$ will be the subject of \cite{usinsecond}.
 
%The result is compatible with the result in \cite{cierischapo}, where it was found that the DQC holds to every order of a perturbative expansion in $\theta$ for a $U_{\star}(2)$ noncommutative gauge theory with a scalar field in the
%BPS-limit, i.e. the minimum energy for the monopole configuration.
%, {\color{magenta} i.e. in the asymptotic limit where plop}

%Furthermore, in \cite{GrossNek} it was found that in the BPS-limit a $U_{\star}(1)$ gauge theory with a scalar field has a soliton solution. In that work the solution is not perturbative,
%but it is neither a magnetic monopole when $\theta$ is turned on. It reduces to the Dirac magnetic monopole when $\theta$ is set to zero, but has zero magnetic charge when $\theta$ is turned %on. Therefore, one cannot say
%much about the DQC in that model. Our result can also be compatible with this result, provided it does not produce a DQC when the expansion is taken to higher orders. 

We would additionally like to mention the fact that although noncommutative QED is known to be CPT invariant \cite{CPT} (see also \cite{Spinstat}), symmetry arguments alone do not rule out the existence of a first order $\theta$-correction term in the DQC.

\noindent {\large \bf Acknowledgments}

We are indebted to Masud Chaichian for illuminating discussions and comments on the manuscript. We also wish to thank Claus Montonen for useful remarks. The support of the Academy of Finland
under the projects no. 136539 and 140886 is gratefully acknowledged.

\renewcommand{\baselinestretch}{1}\selectfont


\begin{thebibliography}{99}

\bibitem{Dirac}
  P.~A.~M.~Dirac,
  %``Quantised singularities in the electromagnetic field,''
  Proc.\ Roy.\ Soc.\ Lond.\  {\bf A 133} (1931) 60.
  %%CITATION = PRSLA,A133,60;%%

\bibitem{WuYa}
  T.~T.~Wu and C.~N.~Yang,
  %``Concept of nonintegrable phase factors and global formulation of gauge
  %fields,''
  Phys.\ Rev.\  {\bf D 12} (1975) 3845.
  %%CITATION = PHRVA,D12,3845;%%

%\bibitem{gia}
 % G. Giacomelli and L. Patrizii, \emph{Magnetic Monopole Searches}, Bologna preprint
 % DFUB-2003-1 (2003). In \emph{Trieste 2002, Astroparticle physics and
 % cosmology} (ICTP, Trieste 2003) p. 121; \\
 % K. A. Milton, Rept.\ Prog.\ Phys.\ {\bf 69}, 1637 (2006).

\bibitem{SeiWitt}
  N.~Seiberg and E.~Witten,
  %``String theory and noncommutative geometry,''
  JHEP {\bf 9909} (1999) 032,
  [arXiv:hep-th/9908142].
  %%CITATION = JHEPA,9909,032;%%

\bibitem{DopFredRob}
  S.~Doplicher, K.~Fredenhagen and J.~E.~Roberts,
  %``Space-time quantization induced by classical gravity,''
  Phys.\ Lett.\  {\bf B 331}, (1994) 39; \\
  %%CITATION = PHLTA,B331,39;%%
  S.~Doplicher, K.~Fredenhagen and J.~E.~Roberts,
  %``The Quantum structure of space-time at the Planck scale and quantum
  %fields,''
  Commun.\ Math.\ Phys.\  {\bf 172}, (1995) 187,
  [arXiv:hep-th/0303037].
  %%CITATION = CMPHA,172,187;%%

\bibitem{GM}
J. Gomis and T. Mehen,
% {\it  Space-time noncommutative field theories and unitarity,}
Nucl. Phys. {\bf B 591} (2000) 265, [arXiv:hep-th/0005129].

\bibitem{SST}
N. Seiberg, L. Susskind and N. Toumbas,
% {\it  Space-time noncommutativity and causality, }
JHEP {\bf 0006} (2000) 044, [arXiv:hep-th/0005040].

\bibitem{ChaKuNiTu}
  M.~Chaichian, P.~P.~Kulish, K.~Nishijima and A.~Tureanu,
  %``On a Lorentz-invariant interpretation of noncommutative space-time and  its
  %implications on noncommutative QFT,''
  Phys.\ Lett.\  {\bf B 604} (2004) 98,
  [arXiv:hep-th/0408069]; \\
  M.~Chaichian, P.~Pre\v{s}najder and A.~Tureanu,
  %``New concept of relativistic invariance in NC space-time: Twisted  Poincare
  %symmetry and its implications,''
  Phys.\ Rev.\ Lett.\  {\bf 94} (2005) 151602,
  [arXiv:hep-th/0409096].
  %%CITATION = PRLTA,94,151602;%%
  %%CITATION = PHLTA,B604,98;%%

  \bibitem{MinRaaSei}
  S.~Minwalla, M.~Van Raamsdonk and N.~Seiberg,
  %``Noncommutative perturbative dynamics,''
  JHEP {\bf 0002} (2000) 020,
  [arXiv:hep-th/9912072].
  %%CITATION = JHEPA,0002,020;%%


\bibitem{Haya}
  M.~Hayakawa,
  %``Perturbative analysis on infrared aspects of noncommutative QED on  R**4,''
  Phys.\ Lett.\  {\bf B 478} (2000) 394,
  [arXiv:hep-th/9912094].
  %%CITATION = PHLTA,B478,394;%%

\bibitem{Gross2}
  D.~J.~Gross, A.~Hashimoto and N.~Itzhaki,
  %``Observables of non-commutative gauge theories,''
  Adv.\ Theor.\ Math.\ Phys.\  {\bf 4} (2000) 893,
  [arXiv:hep-th/0008075].
  %%CITATION = 00203,4,893;%%

%\cite{Shnir:2005xx}
%\bibitem{Shnir}
 % Y.~.M.~Shnir,
 %{\it Magnetic monopoles,}
 % Berlin, Germany: Springer (2005) 532 p.


\bibitem{dqc}
  M.~Chaichian, S.~Ghosh, M.~L\aa ngvik and A.~Tureanu,
  %``Dirac Quantization Condition for Monopole in Noncommutative Space-Time,''
  Phys.\ Rev.\  {\bf D 79} (2009) 125029,
  [arXiv:0902.2453 [hep-th]].
  %%CITATION = PHRVA,D79,125029;%%

%\cite{Jiang:2000wz}
%\bibitem{Jiang}
 % L.~Jiang,
  %``Dirac monopole in non-commutative space,''
 % arXiv:hep-th/0001073.
  %%CITATION = HEP-TH/0001073;%%

\bibitem{U2}
  D.~Bak,
  %``Deformed Nahm equation and a noncommutative BPS monopole,''
  Phys.\ Lett.\  {\bf B 471}, 149 (1999),
  [arXiv:hep-th/9910135]; \\
  K.~Hashimoto, H.~Hata, S.~Moriyama,
  %``Brane configuration from monopole solution in noncommutative superYang-Mills theory,''
  JHEP {\bf 9912}, 021 (1999),
 [arXiv:hep-th/9910196]; \\
   %%CITATION = PHLTA,B471,149;%%
   S.~Goto, H.~Hata,
  %``Noncommutative monopole at the second order in theta,''
  Phys.\ Rev.\  {\bf D 62}, 085022 (2000),
  [arXiv:hep-th/0005101].

\bibitem{U1}
  K.~Hashimoto, T.~Hirayama,
  %``Branes and BPS configurations of noncommutative / commutative gauge theories,''
  Nucl.\ Phys.\  {\bf B 587}, 207 (2000),
  [arXiv:hep-th/0002090]; \\
   L.~Cieri and F.~A.~Schaposnik,
  %``The dyon charge in noncommutative gauge theories,''
  Res.\ Lett.\ Phys.\  {\bf 2008} 890916 (2008),
  [arXiv:0706.0449 [hep-th]].

\bibitem{nonpert}
   D.~J.~Gross and N.~A.~Nekrasov,
  %``Monopoles and strings in noncommutative gauge theory,''
  JHEP {\bf 0007} 034 (2000),
  [arXiv:hep-th/0005204];\\
  %%CITATION = JHEPA,0007,034;%%
   M.~Hamanaka, S.~Terashima,
  %``On exact noncommutative BPS solitons,''
  JHEP {\bf 0103}, 034 (2001),
  [arXiv:hep-th/0010221].
 
\bibitem{lecht} 
  O.~Lechtenfeld, A.~D.~Popov,
  %``Noncommutative monopoles and Riemann-Hilbert problems,''
  JHEP {\bf 0401}, 069 (2004),
  [arXiv:hep-th/0306263].
    %%CITATION = RLPHA,2008,890916;%%
  

\bibitem{usinsecond}
  M.~L\aa ngvik, T.~Salminen and A.~Tureanu, work in progress.

%\cite{SheikhJabbari:2000vi}
\bibitem{CPT}
  M.~M.~Sheikh-Jabbari,
  %``C, P, and T invariance of noncommutative gauge theories,''
  Phys.\ Rev.\ Lett.\  {\bf 84} (2000)  5265,
  [arXiv:hep-th/0001167].

%\cite{Chaichian:2002vw}
\bibitem{Spinstat}
  M.~Chaichian, K.~Nishijima, A.~Tureanu,
  %``Spin statistics and CPT theorems in noncommutative field theory,''
  Phys.\ Lett.\  {\bf B 568}  (2003) 146,
  [arXiv:hep-th/0209008];\\
%\cite{AlvarezGaume:2003mb}
  L.~Alvarez-Gaum\'e and M.~A.~Vazquez-Mozo,
%  {\it General properties of noncommutative field theories,}
  Nucl.\ Phys.\  {\bf B 668} (2003) 293,
  [arXiv:hep-th/0305093];\\
%\cite{Chaichian:2004qk}
  M.~Chaichian, M.~N.~Mnatsakanova, K.~Nishijima, A. Tureanu, Yu.S. Vernov,
  %``Towards an axiomatic formulation of noncommutative quantum field theory,''
  [arXiv:hep-th/0402212].


\end{thebibliography}
\end{document}